\documentstyle[aas2pp4]{article}

\lefthead{Wade \& Hubeny}
\righthead{Detailed UV Spectra for Accretion Disks in CVs}

\received{29 May 1998}
\accepted{24 June 1998}

\begin{document}
\title{Detailed Mid- and Far-Ultraviolet Model Spectra for Accretion
Disks in Cataclysmic Binaries}

\author{Richard A. Wade\altaffilmark{1}}
\affil{Department of Astronomy \& Astrophysics, The Pennsylvania State
University,\\ 525 Davey Laboratory, University Park, PA  16802-6305}
\and
\author{Ivan Hubeny\altaffilmark{1}}
\affil{AURA/NOAO, NASA Goddard Space Flight Center,\\
Code 681, Greenbelt, MD 20771}
\altaffiltext{1}{wade@astro.psu.edu,  hubeny@tlusty.gsfc.nasa.gov}

\begin{abstract} 

We present a large grid of computed far- and mid-ultraviolet spectra
(850 \AA\ to 2000 \AA) of the integrated light from steady-state
accretion disks in luminous cataclysmic variables.  The spectra are
tabulated at 0.25 \AA\ intervals with an adopted FWHM resolution of
1.0 \AA, so they are suitable for use with observed spectra from a
variety of modern space-borne observatories. Twenty-six different
combinations of white dwarf mass $M_{\rm wd}$ and mass accretion rate
$\dot{m}$ are considered, and spectra are presented for six different
disk inclinations $i$.  The disk models are computed self-consistently
in the plane-parallel approximation, assuming LTE and vertical
hydrostatic equilibrium, by solving simultaneously the radiative
transfer, hydrostatic equilibrium, and energy balance equations.
Irradiation from external sources is neglected.  Local spectra of disk
annuli are computed taking into account line transitions from elements
1--28 (H through Ni).  Limb darkening as well as Doppler broadening
and blending of lines are taken into account in computing the
integrated disk spectra.  The radiative properties of the models are
discussed, including the dependence of ultraviolet fluxes and colors
on $M_{\rm wd}$, $\dot{m}$, and $i$.  The appearance of the disk
spectra is illustrated, with regard to changes in the same three
parameters.  Finally, possible future improvements to the present
models and spectra are discussed.  The synthetic spectra are available
as machine-readable ASCII files via {\it ftp\/}.

\end{abstract}


\section{INTRODUCTION}\label{intro}

In most varieties of cataclysmic variable (CV) stars, the mass
transfer between the Roche-lobe filling secondary star and the white
dwarf primary star occurs via an accretion disk.  It is clear from
decades of observations and theoretical studies that this disk can
dominate the ultraviolet and visible spectra of CVs, at least for the
more luminous classes such as novalike variables and dwarf novae in
outburst.  Observations of CVs with the {\it IUE\/} observatory established
the importance of the disk light in the mid-ultraviolet ({\it
mid-UV\/}; roughly, ultraviolet wavelengths longward of H
Lyman-$\alpha$), and {\it Hubble Space Telescope (HST)\/} observations
using {\it HSP\/}, {\it FOS\/}, and {\it GHRS\/} have resulted in
improved data for many key systems, with improved spectral and
temporal resolution and enhanced signal-to-noise ratio.  Observations
with the ultraviolet spectrometers (UVS) aboard the {\it VOYAGER\/}
spacecraft, with the {\it Hopkins Ultraviolet Telescope (HUT)\/}, and
with the Space Shuttle-borne {\it ORFEUS\/} experiments, have shown
the same dominance of the disk in luminous CVs for the far
ultraviolet ({\it far-UV\/}; from the Lyman edge to Lyman-$\alpha$).
The disk may remain dominant in less luminous CVs (mainly the dwarf
novae in quiescence), or it may contribute significantly less than the
white dwarf, especially if the disk is highly inclined to the
observer's line of sight.  In one class of CVs, the AM Her stars, the
accretion flow is channeled by strong magnetic fields and a disk is
not formed.

To make the most of the large and increasing archive of high quality
mid- and far-UV observations of disk-dominated, high-luminosity CVs,
and to distinguish reliably between white dwarf and disk spectra in
less luminous cases, it is clearly necessary to have up-to-date models
for the spectral energy distribution and detailed line spectra of
disks, for a variety of accretion rates, central white dwarf masses,
and viewing angles (orbital inclinations).  As observations have been
collected over the years, different techniques, involving different
levels of approximation, have been applied to this task, and a few
small grids of models have been produced.  For example, Williams and
Ferguson 1982 described a small grid of LTE disk models, spanning the
transition from optically thin to optically thick in the continuum, in
which the disk was assumed to have constant temperature in the
vertical direction at each radius. Wade 1984 discussed the UV and
optical continua for CV accretion disks, comparing two grids of
spectra, based on blackbody energy distributions and computed stellar
fluxes from the Kurucz (1979) grid.  la Dous (1989) introduced some
absorption lines from metals into her computed ultraviolet spectra
of disks, which were LTE models based on the Eddington approximation
without any attempt to ensure energy balance. la Dous also summarizes
prior work by other authors (see also la Dous 1994).

Recent data obtained using the instruments listed above, with their
higher S/N ratios and spectral resolution, demand more detailed
models, computed with some attention given to the properties of disk
spectra that may distinguish them from sums of stellar atmosphere
fluxes. The spectra in Wade (1984), for example, are coarsely binned
and entirely inadequate for the resolution achieved in {\it HUT\/} or
{\it HST\/} spectra. Many studies of individual CVs in recent years
have made use of improved detail in model spectra (e.g., Long et al.\
1994; Knigge et al.\ 1997; Knigge et al.\ 1998), but there has been a
lack until now of a grid of mid- and far-UV spectra of disks,
presented as a body. Such a grid can enable comparisons among objects,
allow the range of temperatures in the disk to be estimated, help
identify lines and blends that can distinguish disk spectra from white
dwarf spectra, and perhaps in favorable cases help determine the mass
accretion rate and orbital inclination in some CVs.

It is also desirable to have a new grid that explicitly tabulates
spectra for disks viewed at a variety of inclinations, since it is
clear that limb darkening of the UV light from a disk is of comparable
importance to purely geometric effects for optically thick disks
viewed at high inclination (e.g. Diaz, Wade, \& Hubeny 1996; Wade
1996; Robinson et al.\ 1995; Robinson, Wood, \& Wade 1998).  This was
shown in the early calculations of Herter et al.\ (1979), and has been
illustrated from time to time (e.g., la Dous 1989), but most
discussions of disk spectra since then have treated angle-averaged
quantities. Diaz et al.\ (1996) discussed several aspects of limb
darkening for a grid of 14 model disks, but tabulated limb-darkened
fluxes at only one wavelength.

In this contribution, we present model mid- and far-UV spectra for
accretion disks in CVs, for a range of white dwarf mass $M_{\rm wd}$, mass
transfer rate $\dot{m}$, and inclination angle $i$, that covers the
expected range of interest for luminous (or ``high-state'') accretion
disk systems. These models treat steady-state accretion disks in
vertical hydrostatic equilibrium, with the energy balance of the disk
computed in a self-consistent way.  The modeling assumptions and
techniques are described in Sections 2 and 3.  Properties of the model
disks and especially their spectra are the subject of Section 4.  Such
grids of model disk spectra need not be perfect to be useful,
especially when used in a comparative way, or as a starting point in
examining which way models should be improved or extended.  We discuss
some of these issues briefly in Section 5.

The spectra discussed here are available in tabular form as ASCII files
via {\it anonymous ftp\/} (see Section 3).

\section{MODEL ASSUMPTIONS AND CHOICE OF GRID}\label{modelgrid}
\subsection{Steady-State Disk Model Computations}

Our grid of spectra is constructed as follows.  The disk is assumed to
be axisymmetric and geometrically thin. The disk is assumed to be in
steady state, that is the mass transfer rate $\dot{m}$ is assumed to
be the same at all radii.  Each elementary area of the visible disk
surface is assumed to make a contribution to the integrated spectrum
of the disk, according to its projected surface area and at a
projected Doppler shift, computed assuming circular Kepler orbits
around a central white dwarf star of mass $M_{\rm wd}$ and radius
$R_{\rm wd}$.

The effective temperature, which describes the total radiation energy
flux at the disk surface, is given by (see, e.g., Pringle 1981)
$$ T_{\rm eff}(r) = T_* x^{-3/4} (1-x^{-1/2})^{1/4}$$
where $x = r/R_{\rm wd}$, and
$$ \sigma T_*^4 \equiv {3GM_{\rm wd} \dot{m} \over 8\pi R^3_{\rm wd}} $$
or 
$$ T_* = 64800~{\rm K} \left[ \left( {M_{\rm wd} \over 1~{\rm M_\odot}} \right)
\left( {\dot{m} \over 10^{-9}~{\rm M_\odot~yr^{-1}}} \right)
\left( {R_{\rm wd} \over 10^9~{\rm cm}} \right)^{-3} \right]^{1/4}. $$
The maximum effective temperature $T_{\rm max} = 0.488 T_*$ occurs
at $x = 1.36$.

The construction of a disk spectrum proceeds in four stages.  First,
the disk is divided into a set of concentric rings or annuli, with
each annulus behaving as an independent plane-parallel radiating slab.
The vertical structure of each ring is computed, using the program
TLUSDISK (Hubeny 1990a, b), which is a derivative of the stellar
atmosphere program TLUSTY (Hubeny 1988). The computation treats the
entire vertical structure of each ring, from the midplane to the
surface, without any artificial introduction of the notion of a
distinct ``disk atmosphere''.  The disk stratification is computed in
the LTE approximation.  Hydrostatic equilibrium is assumed, with a
height-varying (tidal) gravitational acceleration.  Energy balance is
enforced between radiative losses at the disk surface and heat
generation due to viscosity, the latter distributed throughout the
vertical extent of the disk.  It is assumed that there is no flux
incident on the disk surface from outside.

In the specific calculations presented in this paper, the number of
depths between disk mid-plane and disk surface is 99.  The viscosity
prescription is based on a Reynolds number approach (Lynden-Bell \&
Pringle 1974; {K\v r\'\i \v z} \& Hubeny 1986). A value of $Re = 5000$
is assumed for all radii.  A parameter $\zeta$ introduced by {K\v
r\'\i \v z} \& Hubeny (1986) controls the vertical distribution of
viscous heat input.  In the present work, $\zeta = 2/3$. The
contribution of microturbulence to the gas pressure is assumed to be
zero in the present work.

Second, the spectrum synthesis program SYNSPEC described by Hubeny,
Lanz, \& Jeffery (1994) is used to solve the radiative transfer
equation to compute the local, rest-frame spectrum for each ring of
the disk.  In addition to detailed profiles of the H and He lines, the
spectrum synthesis includes ``metal'' lines up to $Z=28$ (Ni),
dynamically selected from the extensive line list of Kurucz \& Bell
(1995; see also Kurucz 1991).  Typically, many thousands of lines are
included in the rest-frame spectrum of each ring. At this stage, the
wavelength array is irregular and differs for each ring.  In addition
to the angle-averaged Eddington flux $H(\lambda)$, specific
intensities $I(\lambda;\mu)$ at a number of emergent angles $i$ are
computed and stored in this step.

Third, the rest-frame intensities are combined to generate an
integrated disk spectrum, using the program DISKSYN5a as described
more fully in \S 3.  Fourth, the monochromatic fluxes $f_\lambda$ are
convolved with an adopted Gaussian instrumental broadening function
and resampled uniformly in wavelength.

\subsection{Choice of Grid Parameters}

In our framework, a steady-state disk model is specified by choice of
the parameters $M_{\rm wd}$, $R_{\rm wd}$, $\dot{m}$, $Re$, and
$\zeta$.  These are supplemented by data that determine the details of
how the programs TLUSDISK and SYNSPEC compute the atmosphere and
spectra.  This section describes the selection of the main parameters
of the grid.

Disk models have been computed on a grid of $M_{\rm wd}$ and $\dot{m}$
with almost-regular spacing but irregular grid boundaries.  The choice
of $M_{\rm wd}$ is assumed to specify $R_{\rm wd}$ through the
mass-radius relation for cold electron-degenerate matter.  We use an
analytical mass-radius relation from P.\ P.\ Eggleton (private
communication; cf.\ Nauenberg 1972). Table \ref{table-1-wd} lists for
each choice of $M_{\rm wd}$ the corresponding white dwarf radius
$R_{\rm wd}$, surface gravity $\log g$ (in cgs units), Keplerian
orbital speed at the surface $v_{\rm surf}$, and the combination
$M_{\rm wd}/R^3_{\rm wd}$ which enters into the formula for $T_*$
given above.  The values of $\dot{m}$ were chosen to increase by
factors of $\sqrt{10} \approx 3.16$, and the intent was to have
$M_{\rm wd}/R^3_{\rm wd}$ increase by the same ratio from one choice
of $M_{\rm wd}$ to the next, so that the same $T_*$ would apply to
several different combinations of $M_{\rm wd}$ and $\dot{m}$ in the
grid; this was achieved for all except the last step in $M_{\rm wd}$,
where the factor is only 3.01.

\placetable{table-1-wd}

The original grid extended only to $\dot{m} \leq 10^{-9}~{\rm
M_{\odot}~yr^{-1}}$, but was later extended.  Thus the labeling of
models by letters, shown in Table \ref{table-2-models}, is slightly
irregular.  This table gives for each disk model: the white dwarf mass
$M_{\rm wd}$, the mass transfer rate $\dot{m}$, the number of annuli
computed, the maximum effective temperature of the disk (occurring at
$x = r/R_{\rm wd} = 1.36$), the Rosseland optical depth $\tau_c$ to the
disk midplane at that disk radius, the effective temperature and
optical depth for the outermost (last) ring, and the radius of the
disk.

\placetable{table-2-models}

For each disk model, stratified atmospheres and spectra were computed
at a set of radii that increase approximately logarithmically from the
innermost radius, which is always $x = R_{\rm min}/R_{\rm wd} =
1.05$. The spacing of rings was chosen to ensure a sufficiently dense
sampling of $T_{\rm eff}(r)$.  The outermost ring was chosen to that
$T_{\rm eff}(R_{\rm out})$ would be near 10,000~K, since cooler zones
at larger radii contribute in only a minor way to the far- and mid-UV
disk flux.  (This point is discussed further in \S 4.4 below.)  Table
\ref{table-3-rings} gives the radii and effective temperature
structure for a number of representative models. For other models the
dimensionless radii $x$ are the same. The temperature structure of
other models can be estimated closely from the tabulated model that
has nearly the same $T_{\rm eff}$ at $x=1.36$ (see Table
\ref{table-2-models}, or calculated directly from the formulas given
above.

\placetable{table-3-rings}

The operation of program TLUSDISK is controlled by a data file that,
among other things, specifies the details of opacity and equation of
state calculations.  In our models H and He always contributed to the
equation of state; in some models cooler than 12,000~K, the first 30
elements of the periodic table were used.  Hydrogen bound-free
continua arising from the eight lowest energy levels were always
included in the opacity calculation.  Continua from fourteen levels of
He I were included for models hotter than 12,000~K, and eight levels
of He II were also included for models hotter than 25,000~K.  Other
opacity sources included electron scattering and free-free continua,
and the H$^-$ bound-free and free-free continua when applicable. For
models cooler than 16,000~K, the atmosphere computation was carried
out with 53 frequency points defining the continuum; for models
between 16,000~K and 30,000~K, 59 frequency points were used; for
models with $T_{\rm eff} > 30,000$~K, 65 frequency points were
employed.

For each ring of each disk model, the atmosphere calculation was
started with a gray LTE disk model (Hubeny 1990a) in hydrostatic
equilibrium, then iterated to thermal equilibrium.  The convergence
criterion was that the last relative change in any model quantity be
smaller than $10^{-3}$.

\section{SPECTRUM SYNTHESIS FOR RINGS AND DISKS}\label{spectra}

The radiative transfer solution to compute the detailed emergent
spectrum from each ring was done by SYNSPEC at essentially full
resolution.  That is, the maximum spacing between adjacent
wavelengths, which is 0.01 \AA\ for $\lambda < 1300$~\AA\ and 0.02
\AA\ for $\lambda > 1300$~\AA, is nearly small enough to resolve the
thermal Doppler width of metal lines. Microturbulent broadening of
lines was assumed to be zero.  The spacing of computed wavelengths in
SYNSPEC is done to ensure that there is a wavelength point at the
center of each line that contributes to the spectrum, and at
least one wavelength point between each adjacent pair of lines.
Thus the wavelength grid is irregular at this stage of computation.

For each ring, both angle-averaged fluxes $H_\nu$ and specific
intensities $I_\nu$ are computed.  The $I_\nu$ computation is done for
$\mu \equiv \cos i = 1.00, 0.75, 0.50$, and 0.25. $I_\nu$ is converted
to $I_\lambda$ in subsequent computational steps.

To compute the integrated disk spectrum, a summation over rings is
carried out.  Each ring is divided into a large number of azimuthal
sectors, and the contributions are summed with appropriate area
weighting and Doppler shifting. The size of the Doppler shift (due to
projected orbital motion of the gas) and the specific intensity both
vary depending on the inclination $i$ of the disk symmetry axis to the
line of sight.  For arbitrary values of $\mu = \cos i$, the intensity
$I_\lambda(\lambda;\mu)$ is interpolated linearly in $\mu$ between the
pre-computed $I_\lambda(\lambda;\mu)$.  Here
$I_\lambda(\lambda;\mu=0)$ is assumed to be zero.

The resulting disk spectrum is sampled rather finely, with a uniform
sampling interval of 0.05 \AA.  Since the orbital Doppler motions of
the gas in the disk have smeared each line substantially for all but
exactly face-on orientations, this high sampling frequency is
redundant.  The spectra are convolved with an instrumental profile and
sampled with four wavelength points per FWHM.  Except when resolution
is specifically discussed, all of the spectra described here or
available as machine-readable files have been convolved with a Gaussian
profile whose FWHM is 1.0 \AA.

Disk spectra have been computed and tabulated for six inclination
angles $i$.  Table \ref{table-4-angles} provides a handy
correspondence between $\mu$, $i$, and $\sin i$, which is the factor
by which orbital motions in the disk are projected on the line of
sight.  The six angles are sufficient to show the dependence of the
observed spectrum on the inclination, but may not always be sufficient
to allow direct interpolation for the purposes of detailed fits of
data.  Denser sampling of $i$ or computation of synthetic spectra for
specific ``off-grid'' angles is straightforward; interested readers
may contact the first author.

\placetable{table-4-angles}

The disk spectra are presented for a {\it ``non-projected''\/} disk
viewed from a distance of $d = 100$~pc.  The calculation thus amounts
to approximating the flux integral
$$f_\lambda(\lambda; \mu) = {1 \over d^2}
\int_{R_{\rm min}}^{R_{\rm max}} dr \int_0^{2\pi} r d\phi $$
$$
\times \int_0^\infty d\lambda' I_\lambda(\lambda', r, \mu) 
V(r, \phi, \mu) \Delta 
$$
where
$$ \Delta \equiv \delta \left(\lambda - \lambda' 
 \left [1 + {1 \over c} \left({GM_{\rm wd}\over r}\right)^{1/2} 
 \sin i \sin\phi \right]  \right). $$

Here $\delta$ is the Dirac delta-function, so that the
Doppler-shifting function $\Delta$ selects the appropriate rest-frame
wavelength $\lambda'$ that contributes at the observer's wavelength
$\lambda$.  Therefore the spectra do correctly take account of
projected {\it velocities\/}.

The visibility factor $V(r,\phi,\mu)$ in the formula makes allowance
for the presence of an opaque central body (the white dwarf) that
occults a portion of the inner disk.  That is, some of the innermost
sectors do not contribute to the integrated disk spectrum, if they
would be blocked from view by the white dwarf.  This visibility factor
is not the same as the visibility factor for a ``flared'' disk, as
discussed by Robinson et al.\ (1995) or Robinson, Wood, \& Wade
(1998).  In the present work flaring effects are neglected (but see
the discussion in \S 4.1).

It is important to note that the results illustrated in this paper, as
well as the tabulated spectra, {\it do not\/} include a factor $\cos
i$ representing the geometric foreshortening of the (flat) disk,
that would normally appear as a multiplicative factor in the
expression above.  This factor is omitted, so that the full effect of
limb darkening alone can be judged from the figures.  

To be compared with actual data, the present synthetic disk spectra
must be supplemented with a model spectrum of the central object,
itself with a suitably disk-occulted lower hemisphere.  This white
dwarf spectrum is omitted from the results shown here. The size of the
white dwarf's contribution to the total system light is discussed
briefly in \S 4.3.  Nguyen et al.\ (1998) have presented some examples
of synthetic solar-abundance white dwarf spectra for the mid-UV.  A
large grid of theoretical solar-composition white dwarf model
atmospheres has been constructed, and will be published in due time
(Hubeny \& Lanz, in prep.). The interested reader may contact I.H.\
for more information and selected models.

ASCII tables of the synthetic disk spectra are available in
machine-readable form by {\it anonymous ftp\/} from {\tt
ftp.astro.psu.edu} in directory {\tt pub/wade/disks}.  There are 26
different files, each corresponding to one disk model (as listed in
Table \ref{table-2-models}).  Each file contains a single header line,
followed by 4601 data lines.  Each data line contains seven columns:
wavelength followed by non-projected flux $f_\lambda$ in ${\rm
erg~cm^{-2}~s^{-1}~\AA^{-1}}$ for six values of $\mu$.  The initial
and final wavelengths are 850.00 \AA\ and 2000.00 \AA, and the
wavelength spacing is 0.250 \AA.  Table \ref{table-5-zzsample} shows
excerpts from the file {\tt zz.grid} for disk model {\it zz}.
Publications making use of these tables should make bibliographic
citation of the present paper.

\placetable{table-5-zzsample}

\section{SOME PROPERTIES OF THE RING AND DISK SPECTRA}\label{properties}

\subsection{Disk Vertical Structure}

A detailed discussion of the stratification of disk models is left to
another paper.  A couple of issues deserve mention, since they relate
to the reliability of the approximations made in computing the present
spectra.

All the rings in all the models have Rosseland optical depths $\tau_c
\gtrsim 1$, where $\tau_c$ is measured vertically, from the outside of
the disk to the disk midplane.  (Note that the Rosseland depth
referred to is computed using only the continuous opacity sources.)
This high optical depth helps to ensure that: (1) the local
photosphere, defined as a geometrical region where most of the
observed spectrum features originate, i.e., where the monochromatic
optical depths measured from the surface are of the order of unity, is
a relatively thin layer on top of the bulk of the disk, consistent
with the plane-parallel approximation; (2) the local atmosphere can be
considered as being laterally homogeneous, also consistent with the
plane-parallel assumption; and (3) the formation of (weak) absorption
lines is not unduly affected by shear due to differential rotation of
the disk, consistent with the hydrostatic approximation.  In disk
models with lower values of $\dot{m}$ than considered here, some rings
may become optically thin in the continuum, so that the thermal
balance of the bulk of the disk is dominated by line cooling. In that
case, the details of radiative transfer in the lines, such as
turbulent or shear broadening, become overwhelmingly important, and
the strong lines are in emission.  In the present models, it is
expected that such effects are at least not dominant for the computed
absorption spectrum.  The lowest values of the midplane optical depth
are attained in the inner disk, inside the maximum of $T_{\rm
eff}$. They are lowest for models with the lowest $\dot{m}$ at a fixed
$M_{\rm wd}$, or for the highest $M_{\rm wd}$ at a fixed $\dot{m}$,
reaching $\tau_c \approx 1$ for $\dot{m} = 10^{-10.5}~M_\odot~{\rm
yr^{-1}}$ and $M_{\rm wd}=1.210~M_\odot$ (model {\it x\/}) at
$x=1.05$.

The assumption that the photosphere of the disk is planar was made in
combining the ring spectra to make the integrated disk spectra.  As
will be discussed elsewhere, the opening angles of the disk are
typically $\theta \sim h/r < 0.1$ radian.  For the most extreme
inclinations discussed here, differential visibility of the ``front''
and ``back'' halves of the disk, combined with a non-negligible
difference in limb darkening of the front and back halves, should be
taken into account (see, e.g., Robinson et al.\ 1995; Wade 1996;
Robinson, Wood, \& Wade 1988).  However, this is most important in
considering situations in which the disk is effectively resolved, such
as by eclipse mapping, rather than for the integrated light of disks.
Another caution is that the present models do not include any
radiation from an outer disk ``rim''.

\subsection{Properties of Spectra of Individual Rings}

Because the disk atmospheres considered here are optically thick,
there is a fairly close resemblance between the emitted rest-frame
spectrum from some disk annulus and the spectrum from a
(plane-parallel) stellar atmosphere, having the same $T_{\rm eff}$ and
with $\log g$ corresponding to the effective gravity at the
photosphere of the disk.  Small details will be different, since
hydrostatic equilibrium is solved in one case under conditions of
constant gravitational acceleration, and in the other case with a
gravitational acceleration that increases approximately linearly with
distance from the midplane. Another potential difference between the
disk and atmospheric structure arises because the total flux is
constant with depth in the case of stellar atmospheres, but departs
from constancy in disks. On the other hand, both stellar and disk
atmosphere models solved in LTE have similar temperature
stratification: in both cases the local temperature monotonically
decreases outward, so major features of the spectrum, such as
absorption line formation and limb darkening, are shared in common.

(To be more general, the local temperature decreases vertically in
disks, only when the viscosity power-law exponent $\zeta$ is chosen to
be larger than about 0.5 for typical cases. For lower values, i.e.,
when the kinematic viscosity does not decrease sufficiently rapidly
towards the disk surface, one faces the so-called disk thermal
catastrophe [e.g. Shaviv \& Wehrse 1986].  This is a sharp increase of
local temperature in the superficial layers, where energy is still
generated by viscous dissipation, while all radiative cooling
transitions are optically thin and therefore ineffective in carrying
away the generated energy.  With our choice of $\zeta$ this
catastrophe does not occur.)

Figure \ref{jjrings} illustrates the computed rest-frame spectra of
three rings (18, 22, 26) from model {\it jj\/}, for the region 1500 --
1600 \AA.  The effective temperatures of these rings are 29150~K,
19650~K, and 13140~K.  Angle-averaged Eddington fluxes, $\log
H_\lambda$, are shown.  The left panel (a) shows the spectra at the
full computed resolution, while the right panel (b) shows the same
spectra convolved with a Gaussian instrumental profile (FWHM = 0.2 \AA,
corresponding to $39~{\rm km~s^{-1}}$) and sampled every 0.05 \AA.
Even with this very modest instrumental broadening, it is clear that
information is lost.  In a disk spectrum, however, the situation is
even worse: the Keplerian speed of these rings would be $\sim1565,
1180$, and $890~{\rm km~s^{-1}}$, respectively, so the blending of
lines due to a mixture of projection factors $v(r) \sin i \sin \phi$
is quite severe, even at low inclination.  For this reason the disk
spectra discussed below are convolved with a wider Gaussian (FWHM =
1.0 \AA) without significant loss from the point of view of comparison
with observation.  Blending is discussed further, below.

\placefigure{jjrings}

Figure \ref{jjrings} also illustrates the dependence of certain
spectral features on effective temperature.  Notable are the
\ion{Si}{2} lines near 1530 \AA, the \ion{C}{4} doublet at 1550 \AA,
and the group of lines near 1560 \AA\ (mainly \ion{C}{1}, \ion{Si}{1}, and
\ion{Fe}{2}), each of which forms at a characteristic range of temperatures.  
In a disk spectrum, each of these spectroscopic features will be
kinematically broadened by an amount characteristic of its radius in
the disk. The integrated disk spectrum thus encodes, in a complicated
way, the $T_{\rm eff}(r)$ structure of the disk.  Decoding this scrambled
message necessarily requires spectrum synthesis of the kind presented
here.

\subsection{Properties of Integrated Disk Spectra}

Figure \ref{bbresn} shows the far-UV spectrum of a nearly face-on
disk, model {\it bb\/}, viewed at inclination $i = 8\fdg1$
($\mu=0.990$), for three different instrumental resolutions (Gaussian
FWHM = 0.1, 1.0, and 3.5 \AA).  The smallest FWHM corresponds to a
resolution slightly worse than that of the {\it ORFEUS} Berkeley
spectrometer or the {\it FUSE\/} spectrometers, while the largest FWHM
corresponds to the {\it HUT\/} spectrometer.  For this extreme case of
minimal Doppler broadening and blending of lines, it is clear that
improving the FWHM from 3.5 \AA\ to 1.0 \AA\ results in an increase of
usable information, while further improving the FWHM to 0.1 \AA\ does
not result in significant new information about line formation in the
disk.  Indeed the difference between the 0.1 \AA\ and the 1.0 \AA\
cases is barely noticeable at the scale of the figure, in the cores of
the lines.  This shows again that the adopted FWHM = 1.0 \AA\ is a
reasonable choice.

\placefigure{bbresn}

At a fixed distance from the observer, a disk can appear brighter if
it is made hotter at constant size (higher $T_{\rm max}$ from higher
$\dot{m}$), if it is made larger while keeping $T_{\rm max}$ the same
(lower $M_{\rm wd}$, hence larger $R_{\rm wd}$, requiring an increase in
$\dot{m}$), or if it is tilted closer to a face-on orientation
(increased projected area, and diminished limb darkening).  A disk
will generally appear ``bluer'' (relatively more flux at shorter
wavelengths) if $T_{\rm max}$ is higher, or if $\mu$ is larger (limb
darkening is stronger at shorter wavelengths).  If $T_{\rm max}$ and $\mu$
are kept constant along a disk model sequence of changing $M_{\rm wd}$
(such as the sequence {\it x, t, p, k, aa\/}), the effects on the
spectrum are expected to be more subtle, due to different effective
gravities and different orbital speeds (hence different line blending)
for corresponding rings in different models.  Some of these effects
are illustrated below, while others are better studied with the aid of
the machine-readable files.

Figure \ref{fnu_mu750} shows, for a fixed viewing angle $\mu=0.75$, how
the flux at 1455 \AA\ varies with changes in the mass transfer rate,
along sequences of constant white dwarf mass and radius, and along
sequences of constant $T_{\rm max}$ (recall $T_{\rm max} = T_{\rm eff}$ at
$x=1.36$).  At constant $M_{\rm wd}$, increasing $\dot{m}$ raises the
effective temperature of each disk ring, and also raises the fluxes
from the disk as a whole.  At fixed $T_{\rm max}$, {\it decreasing\/}
$M_{\rm wd}$ increases the flux from the disk, since the scaling of
temperature with dimensionless radius $x$ is not changed, but the
linear scale of the disk increases with $R_{\rm wd}$.

\placefigure{fnu_mu750}

It was noted earlier that the model spectra presented here do not
include the white dwarf's contribution to the light (but do account
for occultation of the inner disk by the white dwarf).  As a rough
guide for comparing the white dwarf flux with the disk fluxes shown in
Figure \ref{fnu_mu750}, an (unocculted) white dwarf of radius
$10^9$~cm, viewed from a distance $d=100$~pc, produces a flux at 1455
\AA\ of $\log f_\nu \approx 0.5, 0.8, 1.0, 1.1$, and $1.2$ (mJy) for
effective temperatures of 25000, 30000, 35000, 40000, and 45000~K,
respectively.  Fluxes for other white dwarf radii can be derived by
appropriate scaling.  For an opaque disk with $R_{\rm min} = R_{\rm
wd}$, the projected geometrical area of the white dwarf is reduced by
a factor $(1 + \cos i)/2$ owing to occultation by the disk.  The white
dwarf may be competitive with the disk in the mid-UV, provided it is
sufficiently hot, and either $M_{\rm wd}$ or $\dot{M}$ is sufficiently
small, or if the disk is viewed at a high inclination angle.

Figure \ref{fratios2} shows, also for $\mu=0.750$, the behavior of two
disk flux ratios, $f_\nu(1075~{\rm \AA})/f_\nu(1455~{\rm \AA})$ and
$f_\nu(1455~{\rm \AA})/f_\nu(1945~{\rm \AA})$, as $\dot{m}$ is
varied.  The former ratio relates far-UV and mid-UV flux, while the
latter ratio is a measure of the mid-UV slope of the spectrum.  As
before, solid lines connect models at the same $M_{\rm wd}$, while
dashed lines connect models having the same $T_{\rm max}$.  At $T_{\rm
max}$ increases, the flux ratios (or colors) become ``bluer''.  The
mid-UV slope is relatively insensitive to disk properties, and above
$T_{\rm max} \sim 39,000$~K shows hardly any change.  The
far-UV/mid-UV ratio is more sensitive to $T_{\rm max}$, as expected.
Note that the flux ratios are sensitive to disk inclination, through
the effect of limb darkening, discussed below.

\placefigure{fratios2}

The previous two figures illustrated the behavior of disk fluxes for a
fixed viewing direction, $\mu = 0.75$.  Even apart from effects
resulting from the changing projected area, the flux from a disk will
change with $\mu$ owing to limb darkening.  The colors (i.e., flux
ratios) will change as well, since limb darkening is wavelength
dependent.  These effects are illustrated in Figure \ref{fnu_vs_mu},
which shows $\log f_\nu$ {\it vs\/} $\mu$ for two disks, models {\it
bb} and {\it z\/}, at three wavelengths 1075 \AA, 1455 \AA, and 1945 \AA.

Limb darkening occurs for both stars and disks. Invoking the classical
Eddington-Barbier relation, the specific intensity $I_\nu(\mu) \approx
S_\nu(\tau_\nu=\mu)$, where $S_\nu$ is the source function, given in
LTE by the Planck function, $B_\nu(T)$.  Since the temperature
decreases (vertically) outward, the specific intensities at lower
values of $\mu$ sample lower-temperature regions, thereby producing a
spectral energy distribution that is both redder and dimmer .  In the
case of disks there is no averaging over a hemisphere as there is in
the case of stars, because disks lack spherical symmetry.  The full
effect of limb darkening is therefore observable, and disks appear
very considerably dimmer and reddened when viewed at high inclination.
Diaz, Wade, \& Hubeny (1996) have discussed ultraviolet limb darkening
for disks in detail.

\placefigure{fnu_vs_mu}

Attention is now turned from photometric properties to the appearance
of the actual disk spectra.  Figure \ref{mdot} shows a sequence of
far-UV $f_\lambda$ spectra for disks around a $M_{\rm
wd}=0.80~{M_\odot}$ white dwarf, viewed with $\mu=0.50$, normalized
near 1330 \AA.  The quantity that varies is the mass transfer rate
$\dot{m}$.  The dramatic change in overall spectral slope is due to
the increasing temperature of the disk.  The hydrogen lines, too, show
changes in strength and profile.  Two effects are at work for the
lines: (1) as the disk becomes hotter overall, relatively larger areas
of the disk are formed at high temperatures where the H lines are
weak; (2) as the disk is made hotter, the location in the disk at
which the H lines attain maximum strength moves outwards, toward
lesser orbital speeds and smaller associated Doppler broadening.

The change in line shape due to the change in Doppler broadening is
also noticeable in the weaker lines. For example, near 1135 \AA\ note
how ``absorption'' metamorphoses into ``emission'' at the highest
$\dot{m}$. This is actually a case of separate features combining in
the less luminous models to form a single absorption feature, placed
roughly halfway between their rest wavelengths, but only if the disk
is cool enough overall that these features are formed in the inner
disk where orbital speeds are high. (See also Figure \ref{bbmu}.)

\placefigure{mdot}

Figure \ref{mdot_b} shows the mid-UV spectra for the same sequence of
disk models, again viewed at $\mu = 0.50$.  The normalization of
spectra is now at 1950 \AA.  Hotter disks are ``bluer'' (more flux at
short wavelengths), and the detailed line spectra again show
sensitivity to (1) the effective temperatures attained in the disk and
(2) the velocities characterizing the radii in the disk where those
temperatures are attained.

\placefigure{mdot_b}

Another view of the far-UV spectrum is provided in Figure \ref{bbmu},
where a single disk model, {\it bb\/}, is viewed from a variety of
different lines of sight labeled by $\mu$.  Recall that the geometric
projection factor $\cos i = \mu$ is not included, so the change in
flux levels shown in the figure is due to limb darkening alone.  Limb
darkening is very strong in the ultraviolet, and is sensitive to both
temperature and wavelength.  Also evident is the strong dependence of
the shape and strength of features on inclination angle.  Since
different $\mu$'s correspond to different velocity projection factors,
the pattern of minima and maxima in the disk integrated spectrum
changes markedly from one example to the next. Individual features
appear to broaden, split, merge, or shift their locations. The overall
spectrum becomes smoother when the disk is viewed more nearly edge-on.

\placefigure{bbmu}

Note in Figure \ref{bbmu} how local maxima in the disk flux may appear
at the rest wavelengths of strong features, when the feature has been
split in two by the Doppler effect, such as the 1085 \AA\ line of
\ion{He}{2}, viewed at $\mu=0.500$. Lines formed at high temperature
in the inner disk will be broadened more, at a particular inclination,
than lines formed in the more slowly moving, cool outer disk.  Thus a
single ``broadening convolution'' cannot be applied to a disk
spectrum, as it can be to the spectrum of a single star: spectrum
synthesis is mandatory.

A final exploration of the dependence of disk spectra on temperature
is provided by Figure \ref{logflam}, which shows the region 1300 --
1700 \AA\ for a sequence of disks with increasing $\dot{m}$ onto a
0.80 $M_\odot$ white dwarf.  The viewing angle is nearly face-on in
this diagram, and fluxes are plotted logarithmically to facilitate
comparison of feature strengths in one model with another.  ``Cooler''
disks (models {\it m, n, p\/}) show certain absorption features more
strongly than hot disks, and {\it vice versa\/}.  Thus, the integrated
disk spectra do retain, in some measure, information concerning the
local, rest-frame spectra (cf.\ Figure \ref{jjrings}) of the rings
from which they are constituted.  This information appears in a
diluted and convoluted form, which makes interpretation of the spectra
difficult.  However, it is the most direct connection to the physics
of a disk atmosphere.

\placefigure{logflam}

An important question is whether the ultraviolet spectrum of a
steady-state accretion disk is sufficiently distinctive to enable the
white dwarf mass, mass transfer rate, and inclination to be recovered
uniquely.  The preceding discussion shows that in many cases it will
be possible to make useful distinctions among the models.  However, it
is possible to find pairs of spectra that are very similar.  Figure
\ref{matched} compares two models, {\it cc\/} and {\it v\/} which have
the same value of $T_{\rm max}$, achieved through different
combinations of $M_{\rm wd}$ and $\dot{m}$.  Since $M_{\rm wd}$
differs, there is a difference in the Keplerian orbital speed of gas
in the inner disk, but different viewing angles can be chosen to match
the {\it projected\/} velocities of the disk gas, about
3400~km~s$^{-1}$ in this case.  Likewise, the fluxes from the two
model disks can be scaled to match at some wavelength, as if choosing
the distances from which the disks are viewed.  The result of this
matching, as shown by the Figure, is two spectra that are nearly
indistinguishable.  The remaining differences are due to different
amounts of ``reddening'' from limb darkening and a small difference in
the effective photospheric gravity of the two disks. The differences
are subtle and might pass unnoticed or be attributed to differences in
interstellar reddening, for example.  Yet the difference in $\dot{m}$
is more than a factor of three!

Additional, rather precise, knowledge about an observed system may be
required to break such a degeneracy of model spectra that match the
spectroscopic observations about equally well.  This information might
concern the orbital inclination, the white dwarf mass, or the
distance. The specific match chosen for illustration does not involve
matching extreme values of mass, inclination, or distance, however,
and if encountered in practice would require precision of measurement
better than is usually attained, in order to distinguish the models.

\placefigure{matched}

\subsection{Missing Flux Due to Truncation at $T(r) \approx 10,000$~K}

An important consideration in using these models is to what extent
flux is missing, due to the fact that the spectrum synthesis was
stopped when $T_{\rm eff}$ fell below roughly 10,000~K.  This missing
flux may on the one hand affect the overall ``color'' of the
integrated spectra, and may on the other hand affect the details of
line blending and feature strength.  The expectation is that, although
the outer disk (i.e., $r > R_{\rm out}$ of the present models) is
large in surface area, it is cool enough that the contribution in the
far- and mid-UV spectral region can be neglected.

The missing flux question was investigated quantitatively for seven
models ({\it m, s, t, u, v, dd, jj\/}) representing the whole range of
$T_{\rm max}$ of the disk model grid.  From visual inspection of all
the individual ring spectra of model {\it jj\/}, three wavelength
intervals were selected, each 10 \AA\ wide and centered at 1075 \AA,
1455 \AA, and 1945 \AA.  In these intervals, the line blocking is
relatively small and either consistent or smoothly varying over the
range of temperatures that was modeled. For each spectral interval and
each of the seven disk models, mean fluxes $H_\nu$ and intensities
$I_{\nu}$ were computed, ring by ring.  Plots of $x^2 H_\nu(x)$ and
$x^2 I_{\nu}(x)$ against $\ln x$ ($x = r/R_{\rm wd}$) were
constructed, and these were used to extrapolate $x^2 H_{\nu}(x)$ or
$x^2 I_{\nu}(x)$ smoothly to zero in the outer disk.  Numerical
integrations of $H_{\rm total} = 2\pi \int x H_{\nu}(x) dx = 2\pi \int
x^2 H_{\nu}(x) d(\ln x)$ over the extrapolated disk and over the
actually computed disk limits were then compared. A similar procedure
was used with $I_{\nu}$.

In this way it was estimated that at 1075 \AA, an error is made that
is never larger than 0.2 per cent for any disk model, either for
angle-averaged flux or for specific intensity.  For 1455 \AA, the
maximum error grows to 3.5 per cent (for the coolest overall model,
model {\it m\/}).  At 1945 \AA, the maximum estimated missing flux is
largest, as would be expected, amounting to 9.3 per cent for model
{\it m\/}, but no larger than 3.1 per cent for models as hot or hotter
than model {\it u\/}.  The error in intensity is largest when the disk is
viewed nearly face-on ($\mu = 1$). At large inclination angles limb
darkening, which is stronger at a fixed wavelength for cooler
atmospheres, diminishes the relative importance of the missing outer
disk.

As a further check, angle-averaged fluxes $H_\lambda$ for
solar-abundance {\it stellar\/} atmospheres (Kurucz 1994) were used to
synthesize approximate disk spectra corresponding to models {\it m, s,
t, u, v, dd\/} and {\it jj\/}.  These were truncated at $T_{\rm out}
\approx 10000$~K (as with the self-consistent disk models described
here), and alternatively at $T_{\rm out} \approx 4000$~K. Comparison
of the disk fluxes gives truncation errors directly for these stellar
atmosphere-based model disk spectra.  These errors are very similar to
the truncation errors for the self-consistently computed disk spectra
of this paper, as estimated by the extrapolation method described
above.

\section{DISCUSSION}\label{discuss}

Modern observations of CVs in the ultraviolet are frequently of good
enough quality that they deserve to be compared with model spectra
that are more detailed than models used in the past.  The grid of
model spectra presented here is intended to provide a basis for such
comparisons.  The construction of the spectra involves many decisions,
such as what level of approximation is to be used, or what values or
ranges of parameters are to be considered, so these topics have been
addressed at some length.  Also, the illustrative comparison of models
has been emphasized, to show the dependence of the spectrum on
$\dot{m}$, $\mu$, etc.

Detailed analysis of observed CV spectra is beyond the scope of this
presentation, as it involves a host of additional considerations
pertaining to individual objects. These may include such things as
interstellar reddening and the extent to which the white dwarf mass or
orbital inclination may be constrained from orbital studies.  It is
appropriate, however, to review briefly some issues that may arise
during the comparison of model spectra and observed spectra.  We
emphasize that the models presented here, although advanced by
comparison with past efforts, and certainly suitable for use, also
represent a starting point for improved future models.

The present disk spectra do not include the contribution from the
central white dwarf or its boundary layer.  The boundary layer is
expected to contribute mainly in the extreme ultraviolet, $\lambda <
911$~\AA\ (Wade 1991; Polidan, Mauche, \& Wade 1990).  The white dwarf's
contribution to the far- and mid-UV spectrum can be estimated (see \S
4.3), and can be readily computed and included.

The present disk models and spectra are based on vertical hydrostatic
equilibrium.  Observations show that among luminous CVs, many of the
resonance lines and stronger subordinate lines in the UV exhibit
P-Cygni or blue-shifted absorption profiles, indicating that a wind
exists.  Likely, this wind arises from the disk, rather than from the
central star (Pereyra, Kallman, \& Blondin 1997; Proga, Stone, \& Drew
1998).  The direct observational consequence of the wind is that
several of the more obvious spectral diagnostics cannot be used, that
would otherwise provide information about the properties of the disk.
More subtle problems may also arise, if the weak UV lines are formed
in gas that already participates in the outflow, or if the spectrum is
formed in an extended zone that undermines the plane-parallel (i.e.,
laterally homogeneous) approximation that was used in computing the
models of disk annuli.  Further modeling will be required to address
these questions.

The disk stratifications computed here, and the spectra derived from
them, do not include the effect of irradiation by an external source,
whether it is direct illumination of the disk by the central star,
illumination of the outer disk by the hotter inner disk, or emission
or scattering from a wind.  Such irradiation will be most important
for relatively cool zones of the disk (Hubeny 1990b), for which the
contrast between the color temperature of incident radiation and the
photospheric gas temperature is largest.  Since the far- and mid-UV
spectra of disks are dominated by the inner disk, it is reasonable as
a first step to neglect this complication, provided that the disk is
viewed as a whole.  If spectra from isolated portions of the disk are
considered, irradiation becomes a larger issue, even in the UV.

Finally, within the class of hydrostatic, non-irradiated models there
are some important questions to be further explored, concerning,
e.g., line blanketing, or non-LTE treatment of stratification and
radiative transfer.  These issues are common to all computations of
astrophysical atmospheres, whether stars or disks.  The model spectra
presented here serve as a reference, against which to compare spectra
computed under different assumptions or approximations.

\section{SUMMARY}\label{summary}

Based on the need for detailed models to compare with modern
observations of disks in CVs, we have presented a grid of far- and
mid-UV spectra of steady-state, hydrostatic, non-irradiated accretion disks.
The grid covers a significant range of $M_{\rm wd}$ and $\dot{m}$,
suitable for comparison with novalike variables or dwarf novae in
outburst.  The grid consists of 26 model disks, with spectra computed
for each of them at six inclination angles.  The models were computed
using computer programs TLUSDISK, SYNSPEC, and DISKSYN.  A basic set
of continuum opacities was used in the calculation of vertical
stratification, but a large line list was used in the computation of
the emergent spectra.  The emergent spectra take account of limb
darkening, and also of Doppler shifts due to orbital motion of the gas
in the disk around the central white dwarf.  Occultation of part of
the inner disk by the central star is also taken into account.

The spectra extend from 850 \AA\ to 2000 \AA, suitable for comparison
with {\it IUE\/}, {\it HST\/}, {\it HUT\/}, {\it ORFEUS\/}, and {\it
FUSE\/} spectra. The models extend to cool enough temperatures that
essentially all of the light produced by the disk at the wavelengths
of interest is accounted for.  Machine-readable files accessible via
{\it anonymous ftp\/} provide tabulations of the spectra at 0.25 \AA\
steps with a FWHM resolution of 1.0 \AA.  The spectra are presented in
the form of {\it non-projected\/} fluxes for a disk viewed from a
distance of 100 pc.  For comparison with observations, the disk fluxes
may need to be supplemented with fluxes from the appropriately
occulted central star.

The radiative properties of both individual annuli and full disks are
discussed with illustrative examples, to show how the disk spectra
incorporate information about projected gas velocities and the
variation of $T_{\rm eff}$ with radial distance from the white dwarf.
In the disks considered, which are optically thick at all radii, the
local UV spectra resemble spectra of stars that share the same $T_{\rm
eff}$ and $\log g$.  Fluxes and colors (flux ratios) of the integrated
disk spectra depend on $M_{\rm wd}$ and $\dot{m}$.  Limb darkening is
shown to have a strong effect in the ultraviolet, affecting both the
overall flux level and colors of the disk.  Line broadening and
blending is complex and non-intuitive in these disks.  Synthetic
spectra such as those presented here may be needed to understand the
details of observed spectra, or even in some cases to identify the
lines responsible for a given spectral feature.  Despite trends in the
disk spectra that are evident with varying accretion rate or
inclination, spectra from rather different models may nevertheless
appear quite similar, and additional information about the CV may be
required to decide on a unique, best model.

Several possible improvements or additions to the present set of
models are discussed.  As such models become available it will be of
interest to compare the resulting spectra with those presented here.
In the meantime, the present grid of spectra represents a
self-consistent treatment of both the vertical stratification and
emergent spectral properties of steady-state disks, with sufficient
detail in the computation of the line spectrum to be of use in
analyzing observed spectra of CV disks.

\acknowledgements

Support from NASA grants NAG5-1698, NAG5-2125, NAGW-3171, and
NAG5-3459 is gratefully acknowledged.

\newpage

\figcaption[jjrings.ps]{Rest-frame spectra (surface Eddington flux $H_\lambda$)
for three rings of disk model {\it jj}. These rings, Numbers 18, 22,
and 26, have $T_{\rm eff}$ = 29150~K, 19650~K, and 13140~K
respectively.  Left panel: spectra are shown at full computed
resolution, maximum stepsize = 0.02 \AA. Right panel: spectra are
shown after convolution with a Gaussian instrumental profile, FWHM =
0.2 \AA. Different spectral features form at different characteristic
temperatures. \label{jjrings}}

\figcaption[bbresn.ps]{Non-projected far-UV fluxes from disk model
{\it bb}, viewed at 100 pc distance. The disk is viewed nearly
face-on. Three instrumental resolutions are shown, specified by
Gaussian FWHM = 0.1, 1.0, and 3.5 \AA. FWHM = 3.5 \AA\ roughly
corresponds to the resolution of the {\it HUT\/}, while {\it FUSE\/}
is expected to resolve better than 0.1 \AA.  The disk models do not
include wind lines. \label{bbresn}}

\figcaption[fnu_mu750.ps]{Non-projected flux $f_\nu$, 
averaged over the band 1450 --- 1460 \AA, in mJy.  (One milliJansky 
$= 10^{-26}~{\rm erg~cm^{-2}~s^{-1}~Hz^{-1}}$).  The fluxes refer to
disks viewed at a distance of 100 pc.  Limb darkening and line
blending appropriate to $\mu = \cos i = 0.750$ have been taken into
account. Points corresponding to disk models {\it d, m, x, ff}, and
{\it jj\/} are labeled.  Solid lines connect models with the same
$M_{\rm wd}$ (rightmost curve 0.35 $M_\odot$, leftmost curve 1.21
$M_\odot$). Dashed lines connect models with roughly the same
$T_{\rm max}$ (lowest curve $\sim 16,500$~K, uppermost curve $\sim
92,000$~K). \label{fnu_mu750}}

\figcaption[fratios_2.ps]{Ratios of flux $f_\nu$, for disks viewed 
from an inclination $i = 41\fdg4$ ($\mu = 0.750$). Points
corresponding to disk models {\it d, m, x}, and {\it ff\/} are
labeled.  Solid lines connect models with the same $M_{\rm wd}$ (rightmost
curve 0.35 $M_\odot$, leftmost curve 1.21 $M_\odot$). Dashed lines
connect models with roughly the same $T_{\rm max}$ (lowest curve $\sim
16,500$~K, uppermost curve $\sim 92,000$~K. The color range (at a
fixed value of $\mu$) is narrower for the flux ratio involving longer
wavelengths, as expected. \label{fratios2}}

\figcaption[fnu_vs_mu.ps]{Solid circles: the non-projected
fluxes $f_\nu$ at 1075, 1455, and 1945 \AA\ (top to bottom, at
$\mu=1$) for disk model {\it bb\/} ($M_{\rm wd} = 0.550~{M_\odot},
\dot{m} = 10^{-8.5}~M_\odot~{\rm yr^{-1}}$).  Open circles: the same
for disk model {\it z\/} ($M_{\rm wd} = 1.210~{M_\odot}, \dot{m} =
10^{-9.0}~M_\odot~{\rm yr^{-1}}$).  Model {\it bb\/} has
$T_{\rm max}=39110$~K, while model {\it z\/} has $T_{\rm max}=51520$~K.
Colors (flux ratios) and fluxes clearly depend on $\mu$, in addition to
$M_{\rm wd}$ and $\dot{m}$.  Non-projected flux shows the effect of limb
darkening only; multiply by a factor $\mu = \cos i$ to account for
geometric foreshortening.  One milliJansky (mJy) = $10^{-26}~{\rm
erg~cm^{-2}~s^{-1}~Hz^{-1}}$. \label{fnu_vs_mu}}

\figcaption[mdot.ps]{Far-UV fluxes $f_\lambda$, normalized at 1330
\AA, for a sequence of disk models ({\it m, n, p, q, cc, hh\/}) with
constant white dwarf mass but increasing $\dot{m}$ (bottom to top).
All spectra are shown as seen from a viewing direction $i = 60^\circ$
($\mu = \cos i = 0.50$). Spectra are convolved with a Gaussian
instrumental profile, FWHM = 1.0 \AA. \label{mdot}}

\figcaption[mdot_b.ps]{Mid-UV fluxes $f_\lambda$, normalized
at 1950 \AA, for the same sequence of disk models shown in Fig. 6.
Details are the same as for Figure 6. \label{mdot_b}}

\figcaption[bbmu.ps]{Far-UV, non-projected fluxes $f_\lambda$ from
disk model {\it bb\/}, ($M_{\rm wd} = 0.550~{M_\odot}$, $\dot{m} =
10^{-8.5}~M_\odot~{\rm yr^{-1}}$), for six inclinations, $\mu = 0.99,
0.95, 0.75, 0.50, 0.25$, and $0.15$ (top to bottom).  The $\mu = 0.99$
curve is offset vertically by $0.5 \times 10^{-11}~{\rm
erg~cm^{-2}~s^{-1}~\AA^{-1}}$ and shown with a heavier line for
clarity.  Effects of limb darkening and line blending are apparent.
Fluxes should be multiplied by a factor $\mu = \cos i$ to account for
geometrical foreshortening. \label{bbmu}}

\figcaption[logflam.ps]{Details of the mid-UV spectra of a 
sequence of disk models, showing the characteristic features of disks
as a function of disk temperature $T_{\rm max}$.  The spectra are shown as
seen from a viewing angle of $i = 8\fdg1$ so that line blending is
minimized.  Non-projected fluxes are for a viewing distance of 100
pc. As with all figures in this report, the flux from the white dwarf
is not included. \label{logflam}}

\figcaption[matched.ps]{A comparison of two closely matched disk
spectra.  Each model disk ({\it v, cc\/}) has the same run of $T_{\rm
eff}$ with dimensionless radius $x$, and the viewing angles $\mu =
\cos i$ have been chosen to match approximately the run of projected
velocities $v(x) \sin i$.  Finally, the fluxes have been normalized at
1950 \AA\ (upper panel) or 1330 \AA\ (lower panel).  The spectrum of
model {\it v\/} is plotted with a heavier line for clarity.  Residual
differences between the spectra are due to limb darkening and a
systematic difference in effective gravities amounting to about 0.17
dex (see Table \ref{table-2-models}).
\label{matched}}

\clearpage

\begin{deluxetable}{lcccc}
\tablecaption{Adopted White Dwarf Parameters \label{table-1-wd}} 
\tablehead{
\colhead{$M_{wd}$} &   \colhead{$R_{wd}$} & \colhead{$\log g$} &
\colhead{$v_{surf}$} & \colhead{$M_{wd}/R^3_{wd}$\tablenotemark{a}} \\
\colhead{($M_\odot$)} & \colhead{($10^9$~cm)} & \colhead{ } &
\colhead{(km~s$^{-1}$)} & \colhead{ }}
\startdata
0.35 &  1.142 & 7.55  & 2020  &  \phn0.2374     \nl
0.55 &  0.905 & 7.95  & 2840  &  \phn0.7423     \nl
0.80 &  0.699 & 8.34  & 3890  &  \phn2.347\phn  \nl
1.03 &  0.518 & 8.71  & 5140  &  \phn7.423\phn  \nl
1.21 &  0.378 & 9.05  & 6520  &  22.35\phn\phn  \nl
\enddata
\tablenotetext{a}{Units are $M_\odot / (10^9~{\rm cm)^3}$.}
\end{deluxetable}

\clearpage

\begin{deluxetable}{l}
\tablecaption{Parameters of Disk Models \label{table-2-models}}
\tablehead{\colhead{ }} 
\startdata
(EXTERNAL PLANOTABLE, PORTRAIT MODE) \\ 
\enddata
\end{deluxetable}


\begin{deluxetable}{l}
\tablecaption{Temperature Structure of Selected Disk Models \label{table-3-rings}}
\tablehead{\colhead{ }} 
\startdata
(EXTERNAL PLANOTABLE, LANDSCAPE MODE) \\ 
\enddata
\end{deluxetable}

\clearpage

\begin{deluxetable}{ccc}
\tablecaption{Inclination Angles for Computed Disk Spectra\label{table-4-angles}} 
\tablehead{
\colhead{$\mu = \cos i$} &
\colhead{$i$} &
\colhead{$\sin i$}
}
\startdata
0.990 &  8\fdg1 & 0.141 \nl
0.950 & 18\fdg2 & 0.312 \nl
0.750 & 41\fdg4 & 0.661 \nl
0.500 & 60\fdg0 & 0.866 \nl
0.250 & 75\fdg5 & 0.968 \nl
0.150 & 81\fdg4 & 0.989 \nl
\enddata
\end{deluxetable}


\begin{deluxetable}{l}
\tablecaption{Sample Table for Model zz \label{table-5-zzsample}}
\tablehead{\colhead{ }} 
\startdata
(EXTERNAL PLANOTABLE, LANDSCAPE MODE) \\ 
\enddata
\end{deluxetable}

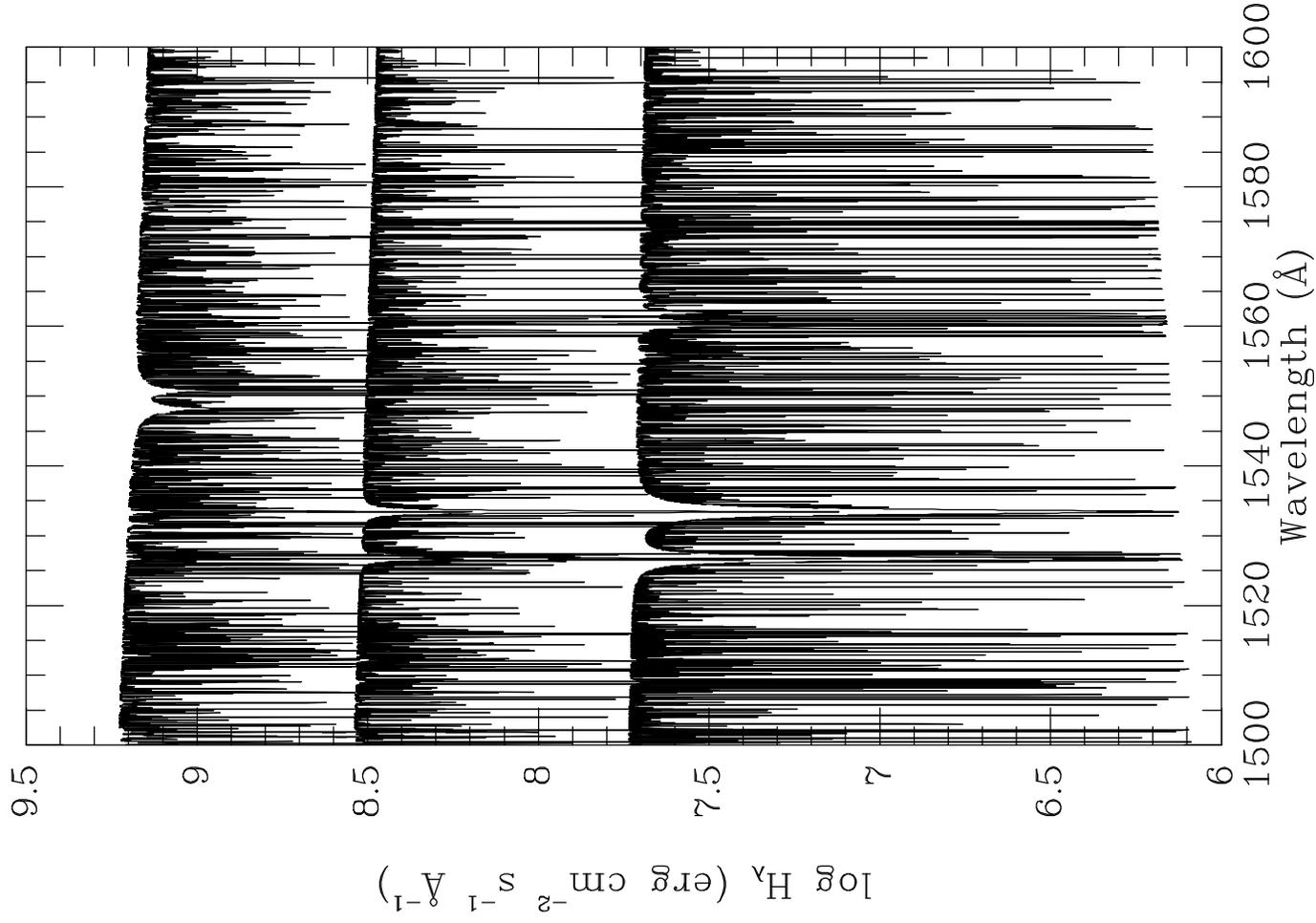
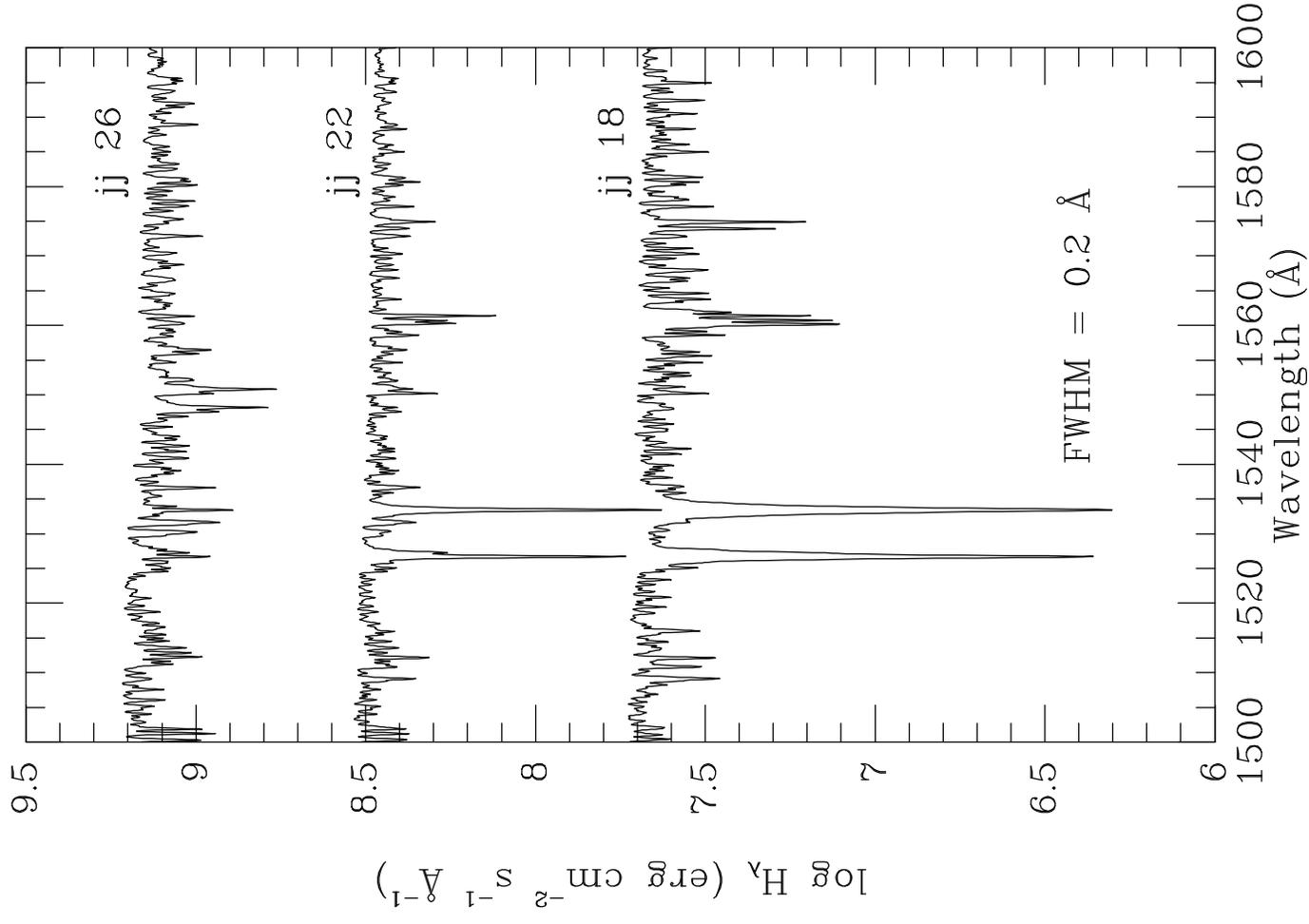
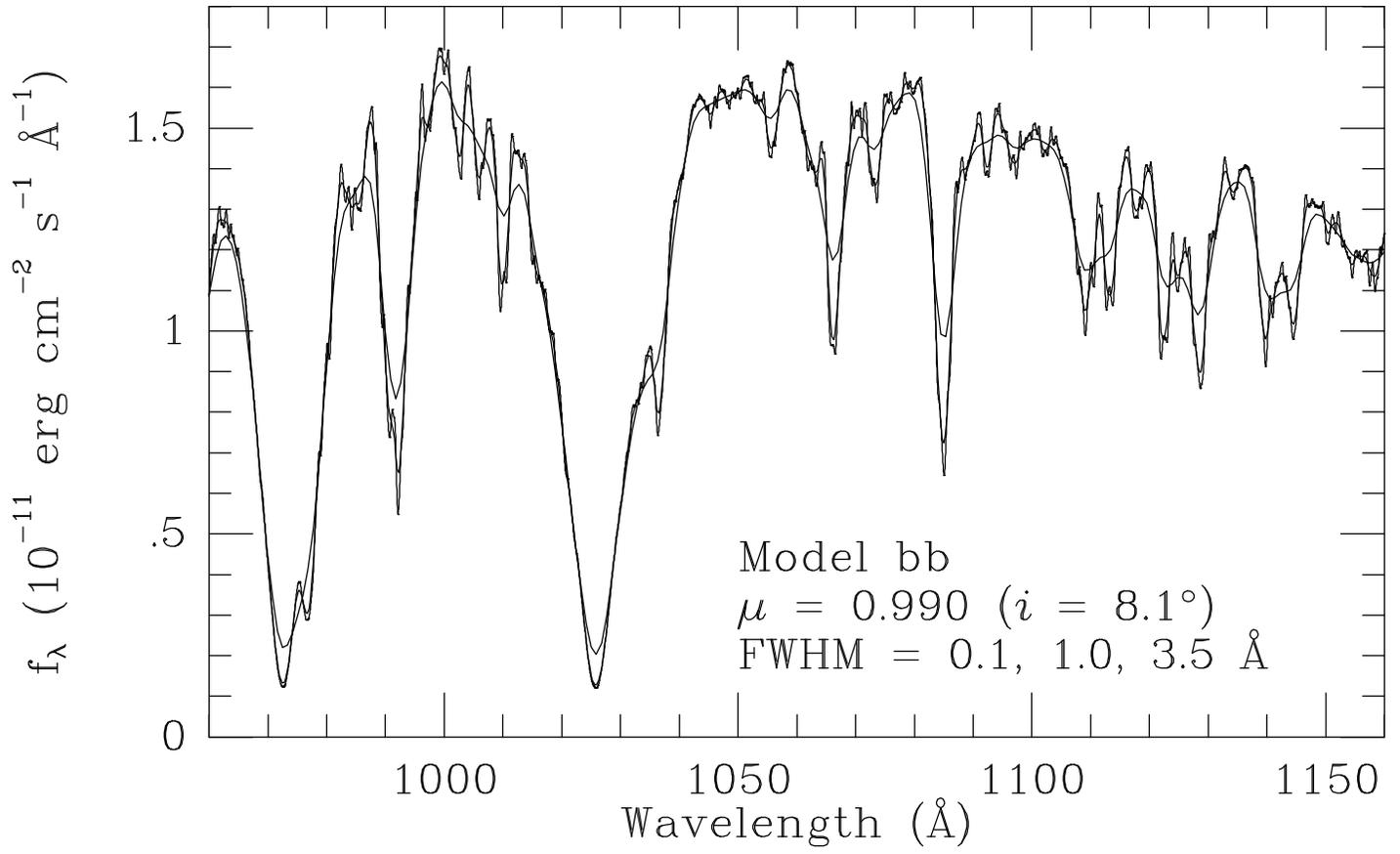
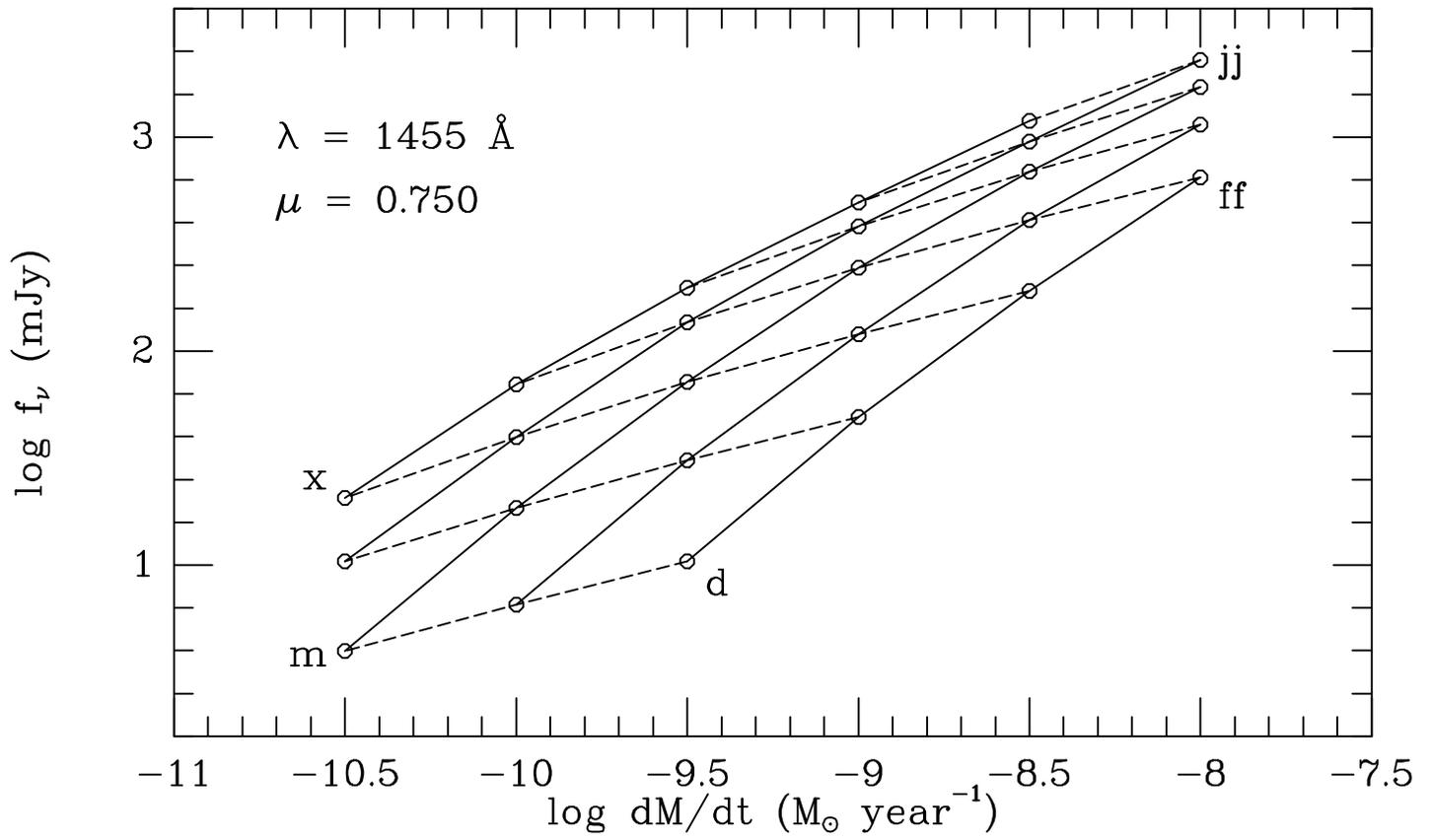
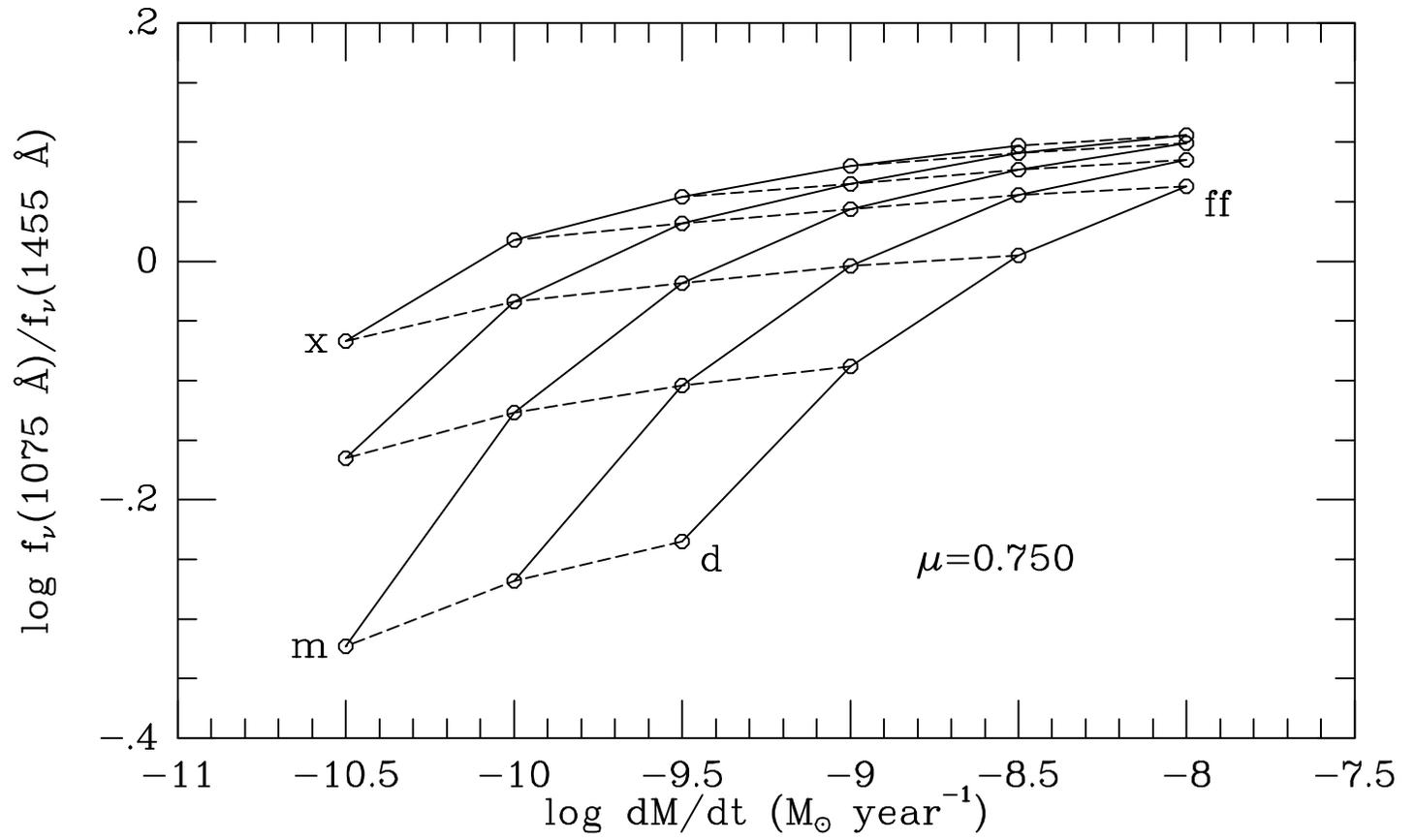
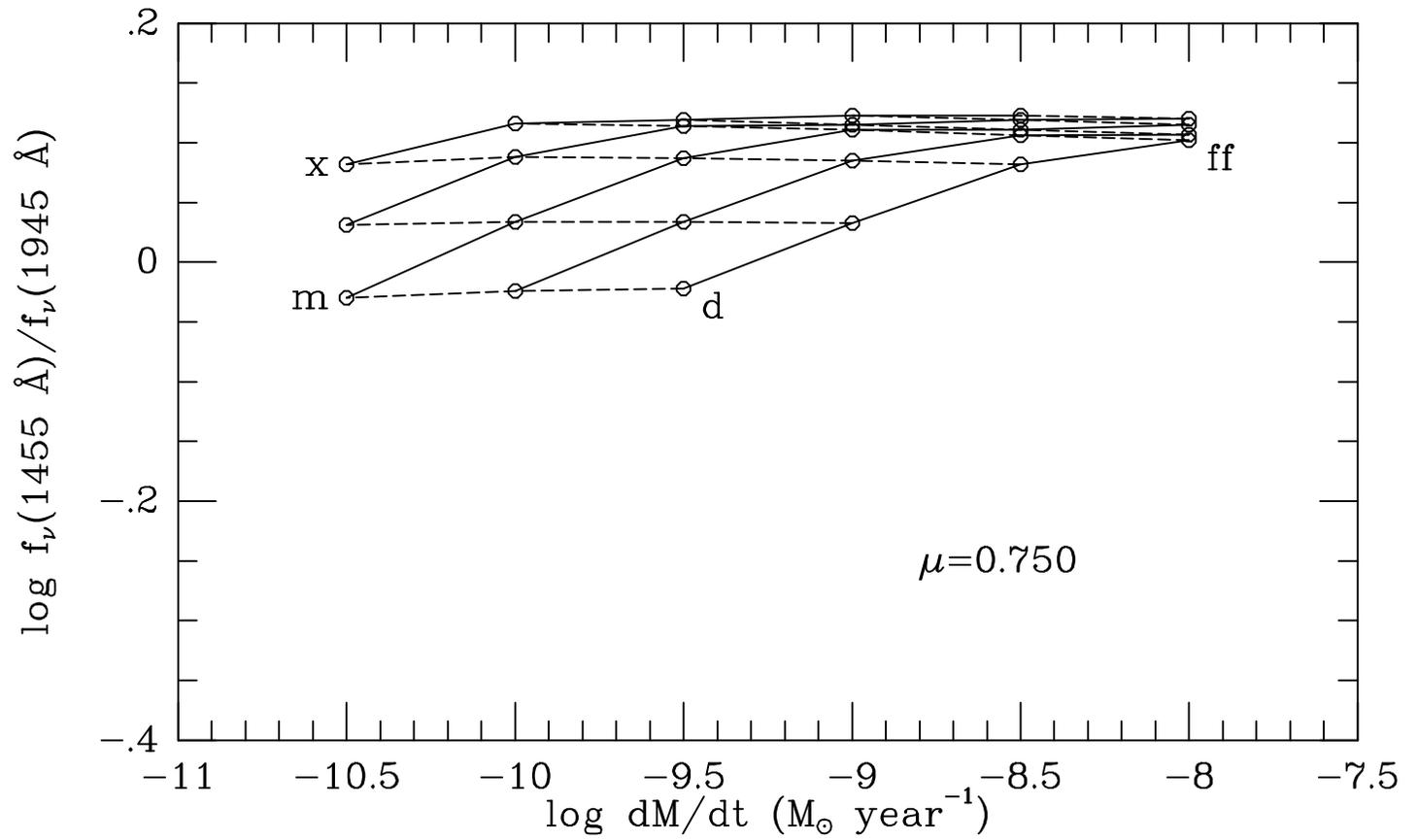
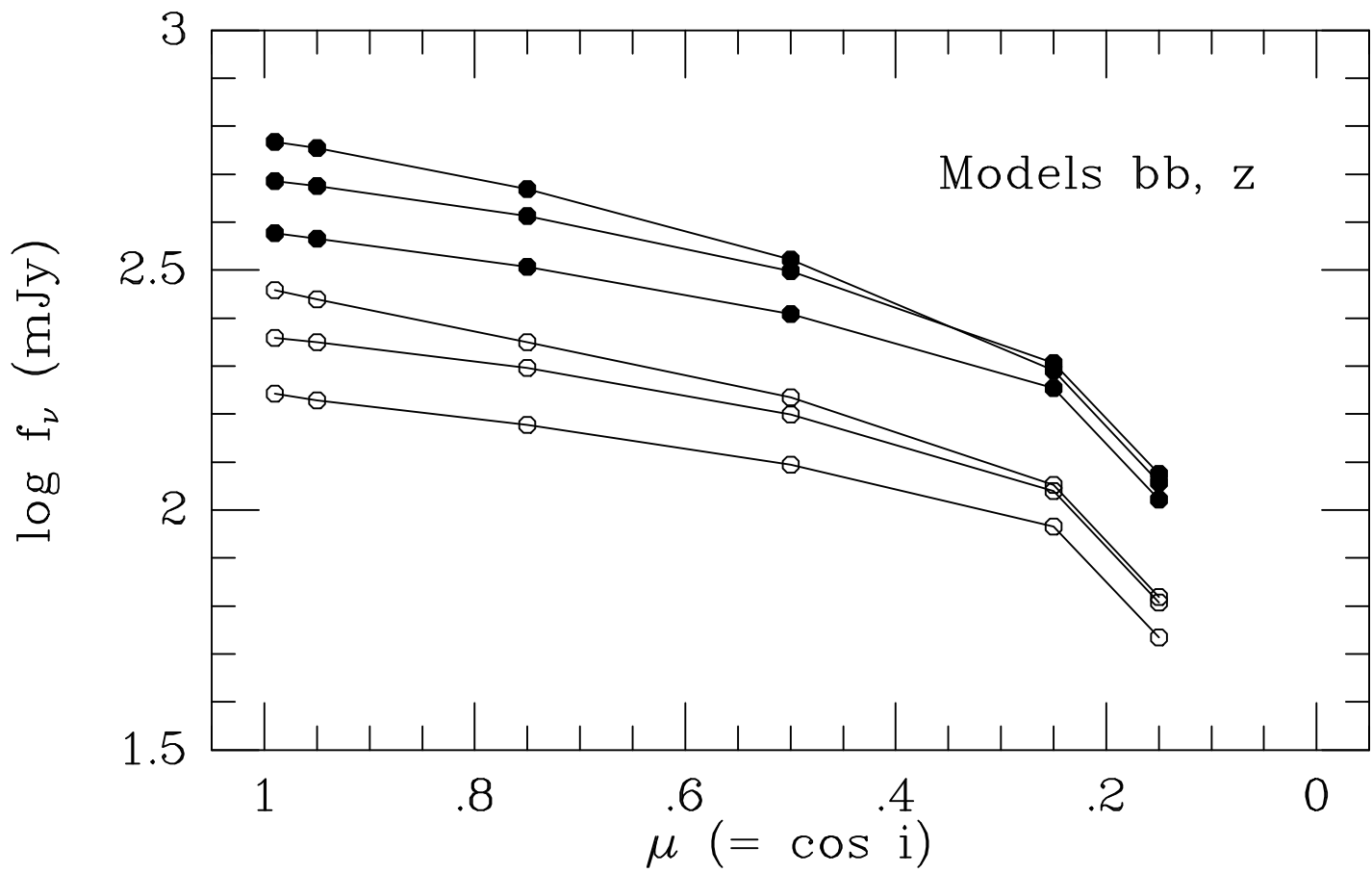
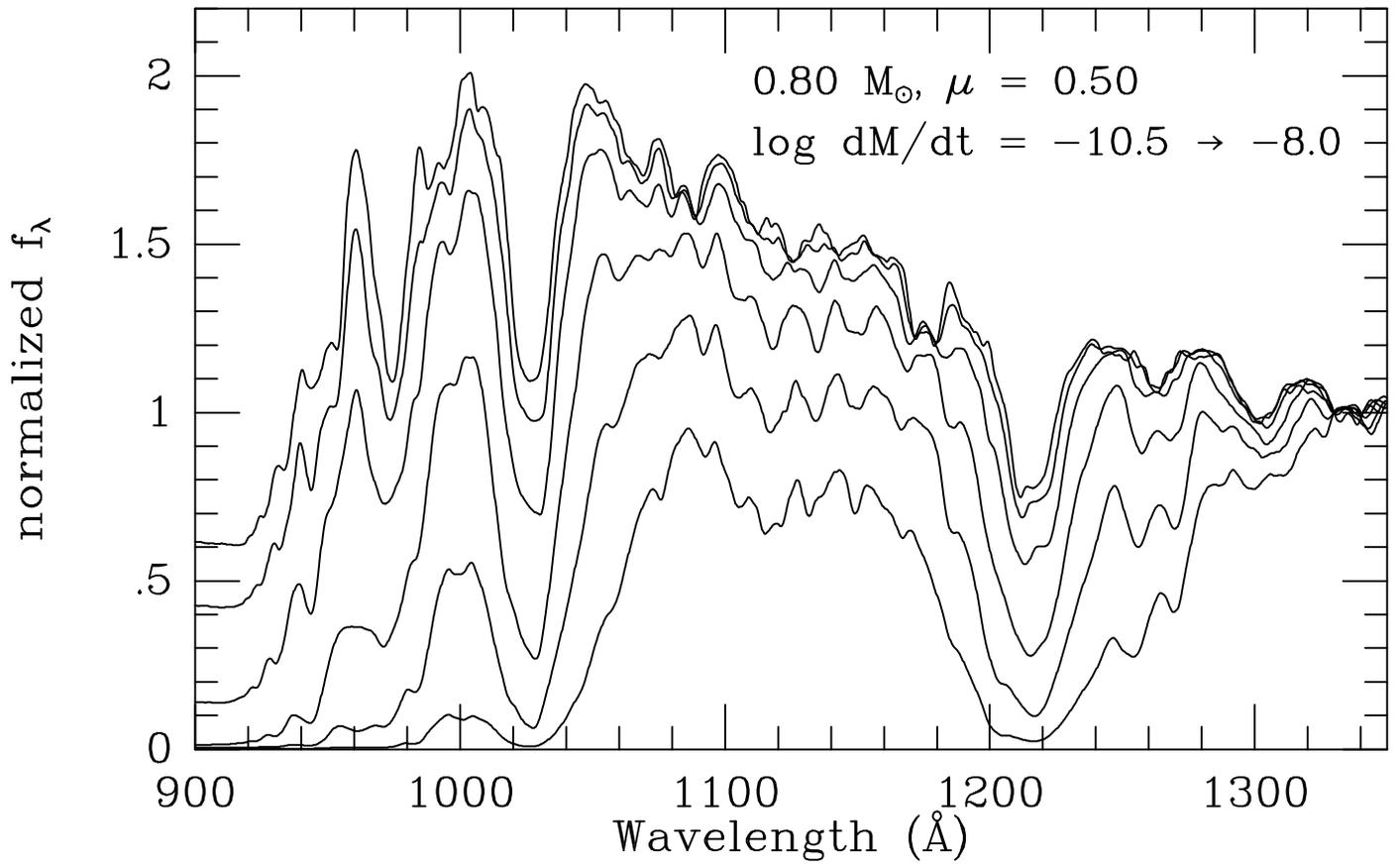
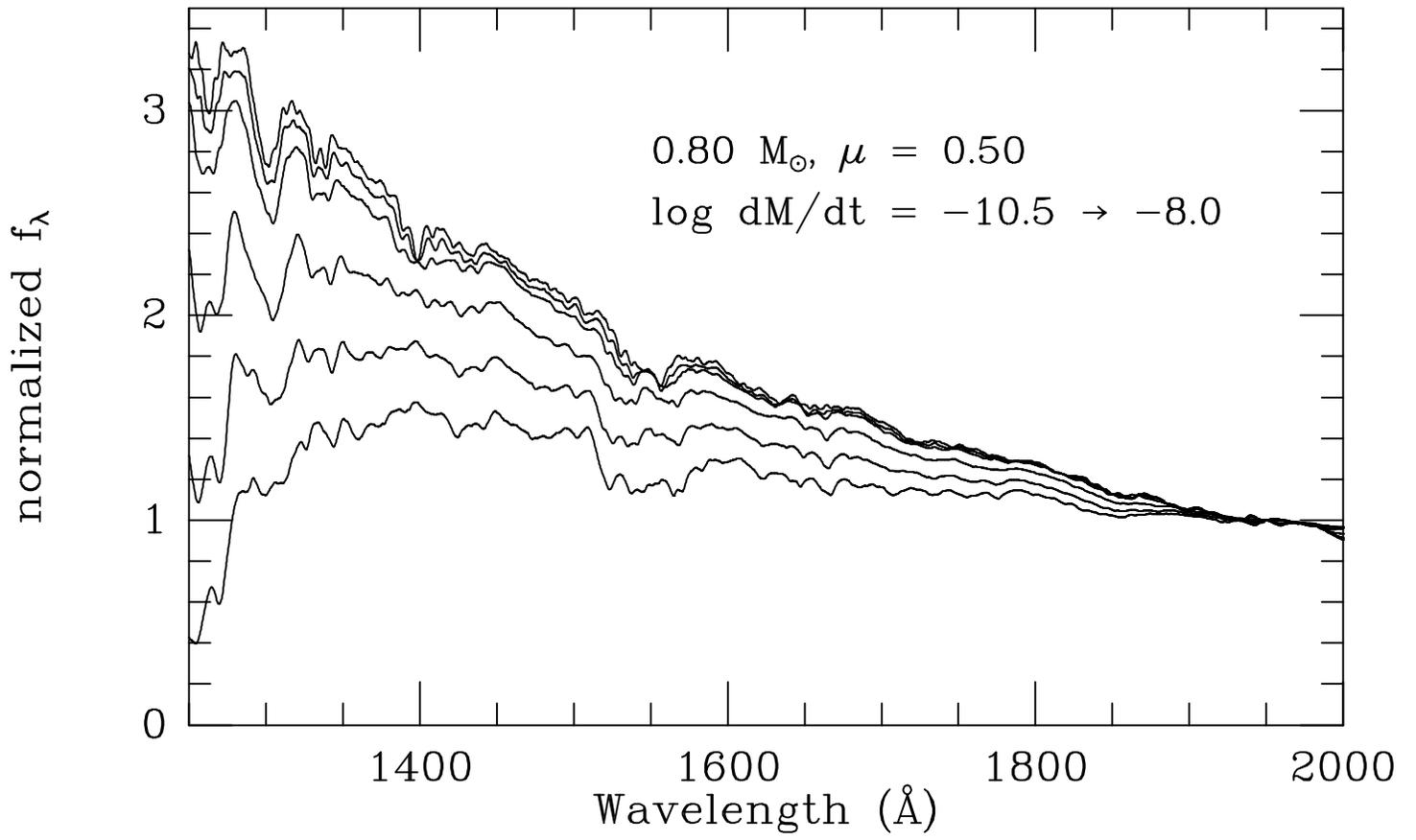
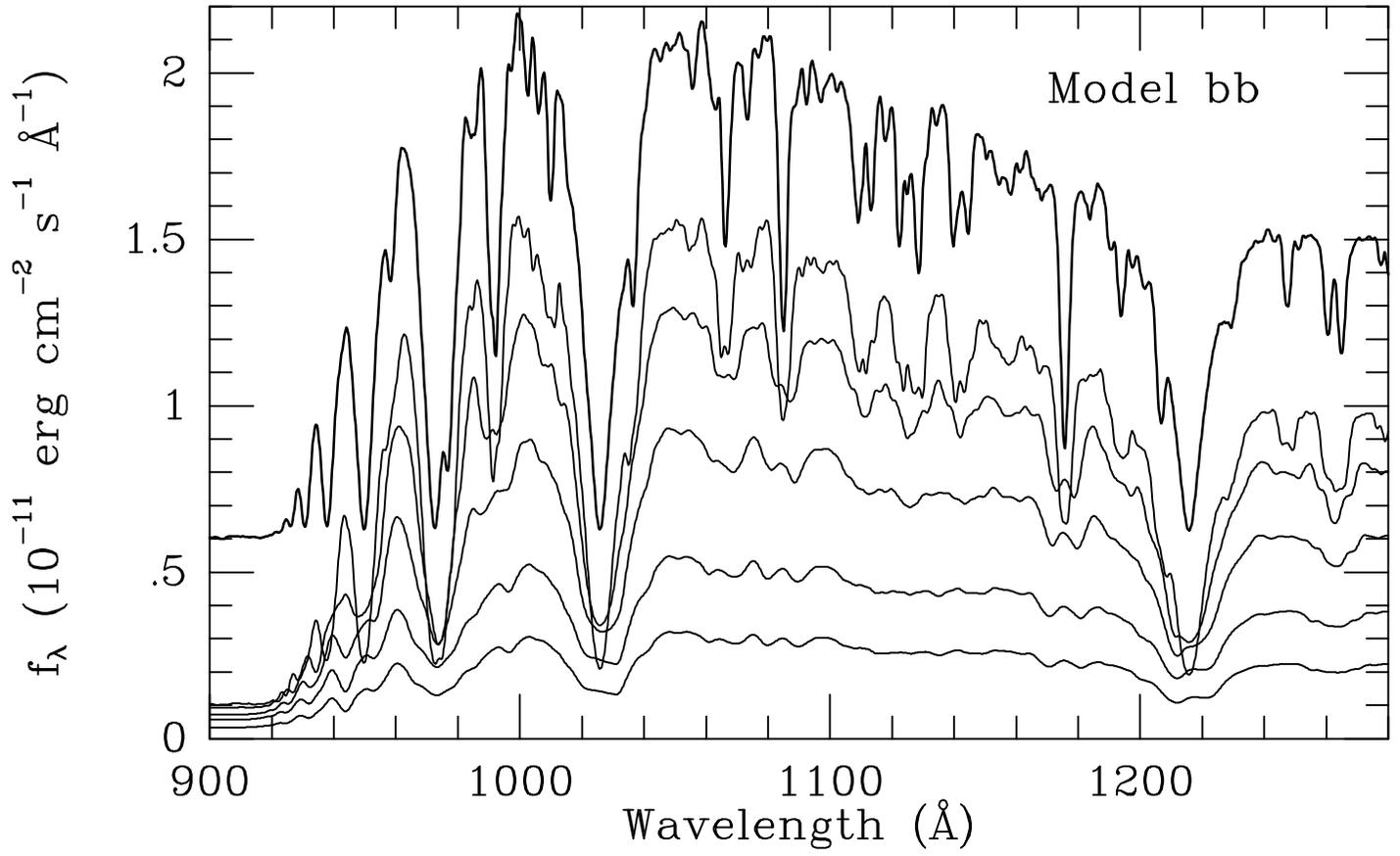
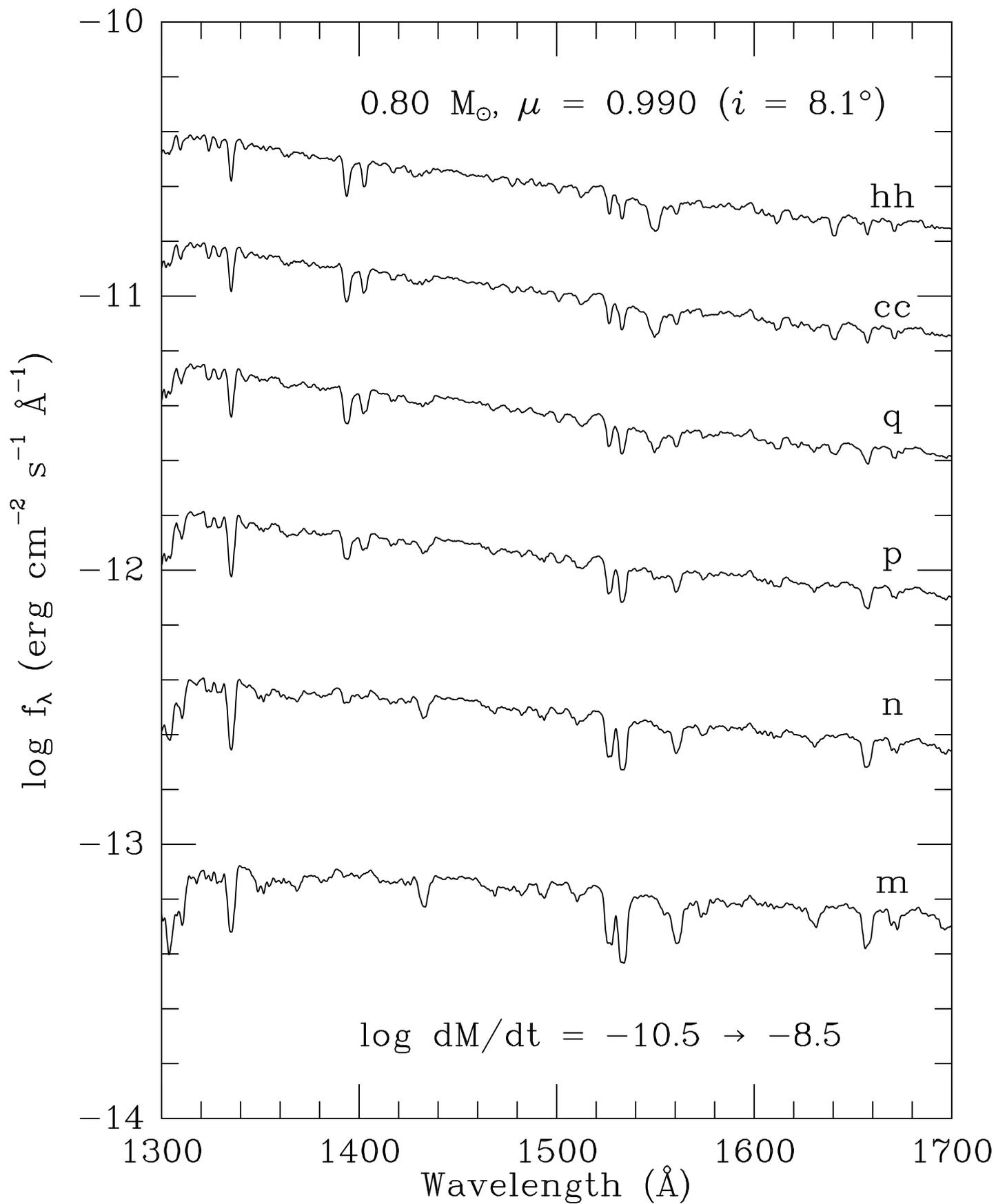
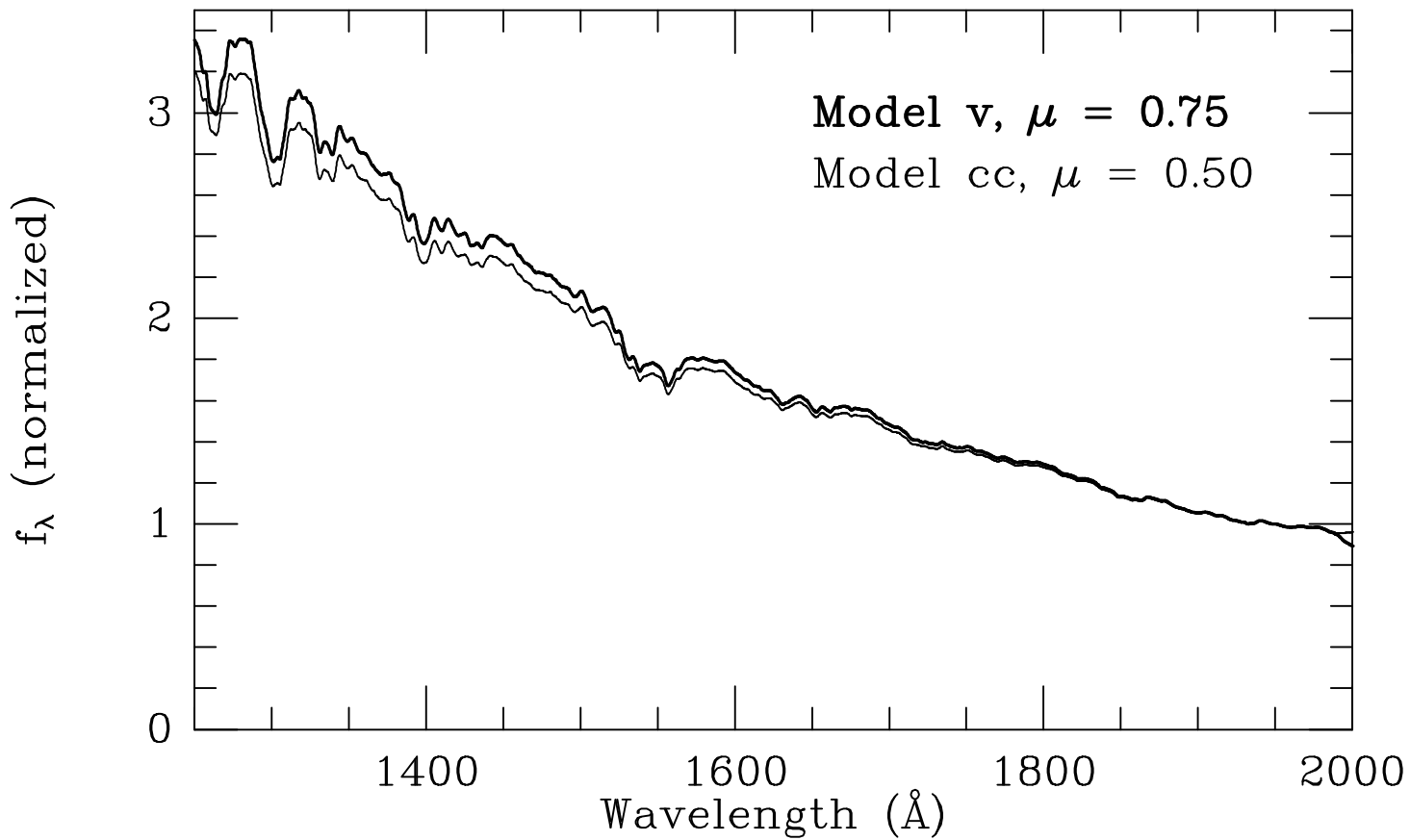
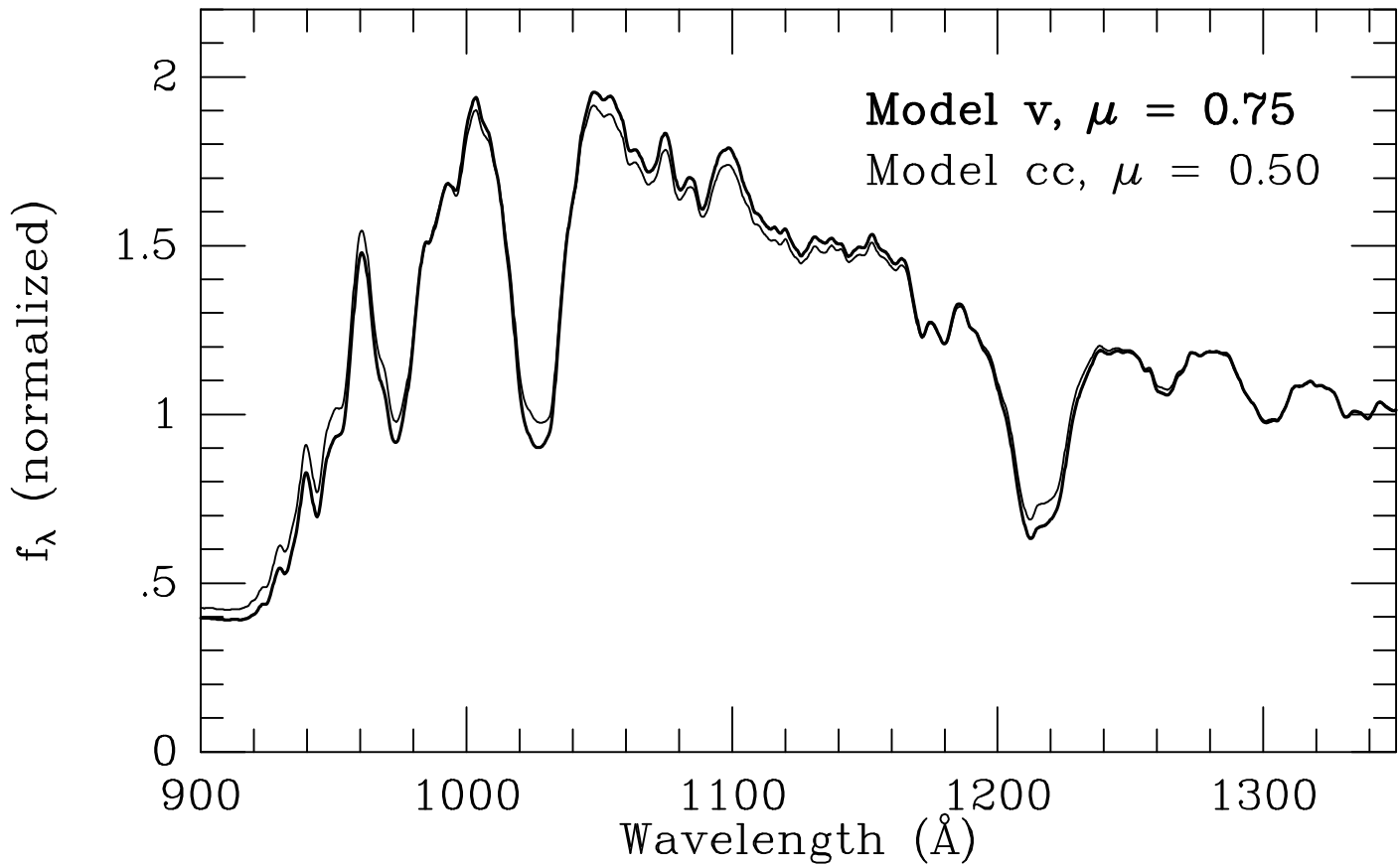
\end{document}